\documentclass[sigconf, 10pt, nonacm]{acmart}

\usepackage{balance}
\let\balance\relax
\renewcommand\footnotetextcopyrightpermission[1]{} 

\usepackage{subfigure}
\usepackage{pifont}
\usepackage[ruled,linesnumbered, vlined]{algorithm2e}
\usepackage{multirow}
\usepackage{enumitem}
\usepackage{colortbl}
\usepackage{tabularx}
\usepackage{bbding}

\usepackage{pifont}
\usepackage{tikz}

\AtBeginDocument{%
  \providecommand\BibTeX{{%
    \normalfont B\kern-0.5em{\scshape i\kern-0.25em b}\kern-0.8em\TeX}}}

\begin{document}

\title[LocaGen]{LocaGen: Sub-Sample Time-Delay Learning for Beam Localization}

\author{Ishaan Kunwar}

\email{kunwar.ishaan@gmail.com}
\author{Henry Cantor}
\email{henrylcantor@icloud.com}
\author{Tyler Rizzo}
\email{tyler.m.rizzo@gmail.com}
\author{Ayaan Qayyum}
\email{aaq18@rutgers.edu}


\begin{abstract}
The goal of LocaGen is to improve the localization performance of audio signals in the 2-D beam localization problem. LocaGen reduces sampling quantization errors through machine learning models trained on realistic synthetic data generated by a simulation. The system increases the accuracy of both direction-of-arrival (DOA) and precise location estimation of an audio beam from an array of three microphones. We demonstrate LocaGen's efficacy on a low-powered embedded system with an increased localization accuracy with a minimal increase in real-time resource usage. LocaGen was demonstrated to reduce DOA error by approximately 67\% even with a microphone array of only 10 kHz in audio processing. 
\end{abstract}

\maketitle

\section{Introduction}
Across engineering disciplines, beam localization plays a key role in ensuring safety, adding to innovation, and supporting a wide range of technologies. In construction, localization is used to find the source of cracks in concrete and preserve integrity \cite{concretePaper}. Localization is even used to implement monitoring in intensive care units \cite{ICUPaper}.

Perhaps one of the most important applications of beam localization is search and rescue (SAR). Natural disasters such as the 2010 earthquake in Haiti result in widespread destruction and mass casualties. Drones have the the flexibility of simpler directional movement, lightweight operations, and the ability to fly over obstacles that may present difficulty for traditional teams. With audio localization technology on drones, an inexpensive device can streamline operations by alerting the operator that a particular area of interest should be searched. As survivors may be immobile and can only call for help, this localization technology can help find victims. Particularly in developing countries, governments and first-responder organizations may not be able to afford traditional SAR drones, as professional drones can cost up to \$30,000 \cite{costSource}. Additionally, visual detection has limited success due to victims being buried under rubble or trapped in collapsed buildings \cite{haitiEarthquake}. System that incorporates audial sensory data to locate victims are instrumental to SAR procedures.

A problem with drone audio and beam localization technology is the need for a compact system. As traditional systems require the detection microphones to be relatively high quality and far apart from each other, the effective size and weight of the device for proper operation is significant. These modules with large footprints require larger, often custom-built drones to support, Such drones may require licenses to operate in some countries, limiting their effective use. This may explain the implementation of commercial drones primarily for visual identification, rather than audial sensory uses, with regard to SAR operations 

Utilizing the performance boost provided by LocaGen, the distance between microphones can be reduced, allowing for lighter, more flexible drones to localize distress signals in SAR operations. The system is trained to localization signals in a 2-D plane. Although the precise angle and distance from the drone to the beam can be estimated, LocaGen does not estimate the relative elevation between the drone and the beam. While the system can be extended to localize a distress signal in 3 dimensions to find the precise elevation of the origin beam, more microphones are needed, leading to larger weight and design complexities. 

\section{Motivation and Preliminaries}
\subsection{The Audio Localization Problem}
This section will explain the problem statement and give an explanation of the variables that went into deriving the number of samples of difference. 

\subsubsection{Explanation of GCC-PHAT}
The generalized cross-correlation phase transform (GCC-PHAT) algorithm computes the cross-correlation between multiple audio signal input channels. The goal of the algorithm is to estimate the time delays between a delayed signal and a reference signal. It finds the estimated delays and cross correlations between all pairs of channels. By examining all pair of microphone signals, and given that the distances between each microphone is specified to a high precision beforehand, the precise estimated time delay can be used to estimate the original location of the sound beam. 

\begin{figure}[h!]
    \centering
    \includegraphics[width=0.2\textwidth]{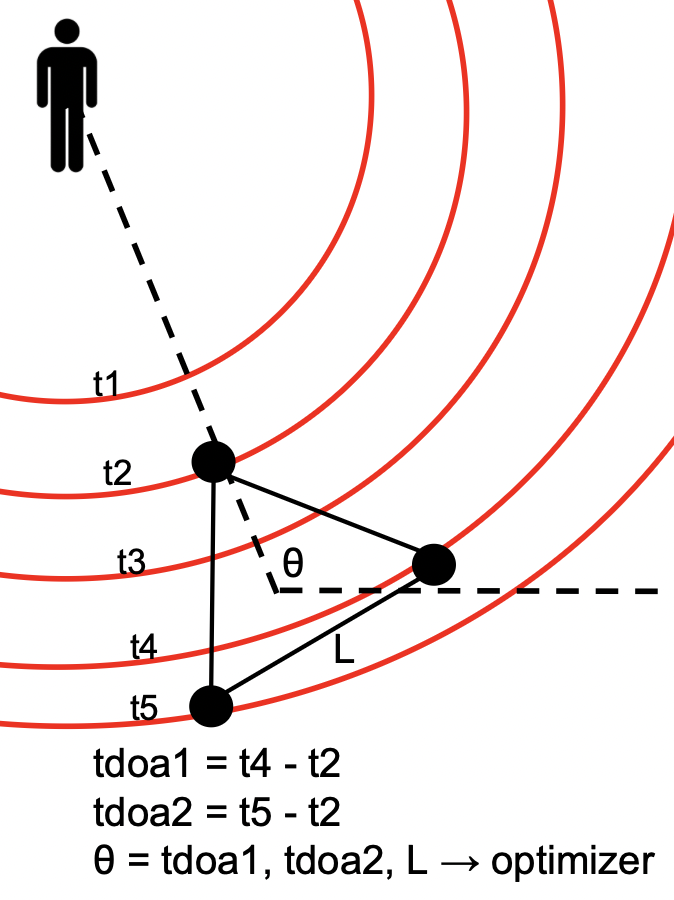}
    \caption{\textbf{Localization Diagram} \\
    The angular location of the source of the scream is determined by the differences in time of arrival and triangulated with three microphones.}
    \Description{The angular location of the source of the scream is determined by the differences in time of arrival and triangulated with three microphones.}
    \label{fig:loc-diagram}
\end{figure}

The GCC-PHAT algorithm takes two signals, $x_i(t)$ and $x_j(t)$, for input. Let $x_i(t)$ and $x_j(t)$ be the time-domain signals captured by two microphones $i$ and $j$, where $x_j(t)$ is a delayed version of $x_i(t)$ due to the geometry of the sound source and the microphone array. Therefore:

\[
x_j(t) = x_i(t - \tau_{ij}) + n(t)
\]

Here, $\tau_{ij}$ is the unknown time delay between the two signals, and $n(t)$ is additive noise.

The signals are transformed into the frequency domain using the Short-Time Fourier Transform (STFT) or standard Fourier Transform:

\[
X_i(f) = \mathcal{F}\{x_i(t)\}, \quad X_j(f) = \mathcal{F}\{x_j(t)\}
\]

The cross-power spectral density is computed as:

\[
R_{ij}(f) = X_i(f) \cdot X_j^*(f)
\]

The complex conjugate of $X_j(f)$ is represented above as $X_j^*(f)$. This captures both the magnitude and phase differences between the two signals.

The necessity to emphasize the phase and suppress the magnitude emerges from the effect of noise or gain corrupting the magnitude result. The Phase Transform (PHAT) weighting is applied for this purpose.  

\[
R_{ij}^{\text{PHAT}}(f) = \frac{R_{ij}(f)}{|R_{ij}(f)|} = \frac{X_i(f) \cdot X_j^*(f)}{|X_i(f) \cdot X_j^*(f)|}
\]

The inverse Fourier transform is used to compute the time-domain cross-correlation function:

\[
r_{ij}(\tau) = \mathcal{F}^{-1}\left\{R_{ij}^{\text{PHAT}}(f)\right\}
\]

Thus yielding a function $r_{ij}(\tau)$ whose peak indicates the estimated time delay $\hat{\tau}_{ij}$ between the two channels.

\[
\hat{\tau}_{ij} = \arg\max_{\tau} \, r_{ij}(\tau)
\]

Given that the relative positions of the microphones are known (and calibrated), the estimated TDOAs $\hat{\tau}_{ij}$ for all microphone pairs $(i, j)$ can be used to geometrically solve for the sound source location.

For example, for a pair of microphones with spacing $d$ and speed of sound $c$, the direction of arrival (DOA) $\theta$ can be computed by:

\[
\theta = \arcsin\left( \frac{c \cdot \hat{\tau}_{ij}}{d} \right)
\]

However, DOA alone is insufficient for determining the full location of the source. Therefore, our system uses a minimum of three microphones. By using multiple delay estimates from different microphone pairs arranged in a known spatial configuration, we can compute the intersection of multiple hyperbolae, each defined by a TDOA measurement. In our system, we aim to estimate the 2D location and angle. This multilateration process enables the estimation of the absolute position of the sound source in space.

Given $N \geq 3$ microphones, we can calculate $\frac{N(N - 1)}{2}$ pairwise time delay estimates $\hat{\tau}_{ij}$, which together constrain the possible positions of the source. Solving this system of geometric constraints yields the estimated source position $\mathbf{s} = (x, y)$ in 2D or $\mathbf{s} = (x, y, z)$ in 3D, depending on the array geometry and available data.

\subsection{Effects of Error in the Real-World}
One of the key problems with determining the location of a beam has to do with the resolution of the microphones themselves. The system should not modify the shape of the signal to increase the accuracy of localization, because there will always be noise in the environment and other imperfections caused by problems with the microphone itself. The goal of LocaGen is to increase the accuracy of the small microphone array to emulate a large microphone array and a higher sampling rate. Therefore, LocaGen introduces an algorithm that tries to correct for quantization errors caused by a low sampling rate and a microphone array that is too close together to make distinctions between samples. 

Of course, errors remain in the beam localization process. For example, data analysis and sending often require compression or feature reduction via quantization. This has the result of reducing localization accuracy.  Additionally, the resolution of microphones in terms of its sample rate constrain the amount of data that can be collected and analyzed to find the source of a sound. Therefore, it is necessary for the system to increase the data points used to train beam localization models and increase the resolution of data collected for high accuracy beam localization.

LocaGen is intended to achieve a result that could have the linear distance between each pair of microphones be only a few centimeters apart yet have the localization and angle estimation accuracy of an array that is only a few meters large. The advancement made by the system is the development of a simulation and data generation toolkit that was used to generate millions of realistic samples and situations to train a machine learning model in reducing angle and location estimation errors. Our simulation is flexible enough to be dynamically adapted to real-life situations.

Although the GCC-PHAT algorithm is intended to mitigate for the effect of sample quantization error by examining the phase information to interpolate between samples, it is limited when it comes to errors caused by the sample rate. With a lower sample rate and the intention of estimating the TDOA and DOA, there is an inherent floor to the precision of each estimation made with the GCC-PHAT algorithm. This floor results from sub-sample jitters and quantization errors. 
\[
\Delta d = v \cdot \frac{1}{f_s} \approx 343 \cdot \frac{1}{48000} \approx 7.1 \, \text{mm}
\]
In this example with the sample rate set to 48 kHz and the speed of sound of 343 m/s, the minimum precision of a distance that can be estimated is limited to around 7.1 mm. As the sample rate goes to infinity, the imprecision caused by quantization error goes to 0. Of course, increasing the size of the microphone array is helpful by allowing more samples between each microphone to occur. However, due to the precision of each microphone's sampling rate, errors are bound to occur in the timescales of a sub-sample. 

In many real-world scenarios, it is impractical to use highly precise equipment due to cost, size, weight, or other restraints. Instead, what may be more practical is the use of interpolation to recreate the effect of an infinite sample rate. With the simulation, the ground truth of the distance can directly be compared to the result of running the real system with the limited sample rate. With the result of millions of localization attempts specifically tuned for a particular microphone array, a machine learning model can be trained to reduce the effect of quantization error to near-zero. However, the challenge of obtaining this data remains a challenge, particularly due to the need to test various different frequencies and distances. 

\subsection{Key Insight: Synthetic Data Generation}
The primary innovation of the LocaGen system was to create a highly automated and scalable system to generate high-quality synthetic data that creates a digital twin of the real-world microphone array. With the design of LocaGen, millions of localization attempts can be directly compared to the true location of the simulated beam. 

\section{LocaGen System Design}

\subsection{System Overview}
\begin{figure}[h!]
    \centering
    \includegraphics[width=0.5\textwidth]{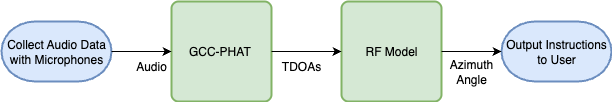}
    \caption{\textbf{Pipeline Flow} \\
    The diagram demonstrates the sequential application of a random forest machine learning model following the GCC-PHAT calculation, in order to output an estimated DOA value.}
    \Description{The diagram demonstrates the sequential application of a random forest machine learning model following the GCC-PHAT calculation, in order to output an estimated DOA value.}
    \label{fig:loc-flowchart}
\end{figure}

\subsection{Simulation Framework}
Our simulation isolates three key parameters, namely the sampling rate, the spacing between microphones, and the speed of sound. We arranged the microphones at the vertices of an equilateral triangle of side length $d$. 

The sampling rate (Hz) limits the temporal resolution of the time-delay estimate. A series of sample rates were tested, including 10 kHz, 24 kHz, 44.1 kHz, and 48 kHz. The microphones used in our system in real life have a maximum sample rate of 48 kHz, which we downsampled to 10 kHz to keep computational costs low. 

In our simulated environment setup, we first define the key parameters that influence the final estimated angle and position. These parameters include the microphone positions, the speed of sound, and the sample rate that matches the real-world microphone. Noise can also be added and specified. The fact that the speed of sound varies based on temperature is taken into account. For each synthetically generated sample, the location of a beam source is specified beforehand. Each microphone records the time the wave is reached, as well as the true arrival time. The true arrival time is based on distance and the speed of sound, not including the effect of quantization error. When the microphone records the time that the wave is reached, a microphone inherently takes its measurement of arrival time after the result of sampling. 

When the beam reaches the microphone, the time that is recorded is affected by the microphone's sample rate. The true distance is recorded for comparison later. Each pair of microphone calculates its estimated time distance of arrival as it would in a real-world scenario, saved to the variable X-train. The Y-true variable records the true and accurate time distance of arrival vectors. Each microphone also has artifical noise added that is unique per microphone. In testing, Gaussian noise was used. The goal is to introduce random variability in the collected sample that will reduce the accuracy of the result from the GCC-PHAT algorithm to better account for real-world imprecision. 

Imprecision is also introduced to account for the fact that in the real world, ensuring all pairs of microphones are exactly the specified distance apart to an infinitely precise degree is difficult. Instead, the distance specified between each microphone allows for tolerances up to 0.1 cm, allowing for flexibility and resilience to form the basis of the synthetic data. 

The above simulation runs to collect TDOA pairs of all microphones from as many samples as desired by the user. At this step, the minimizer method is used to identify the azimuth angle that leads to the least Euclidean distance between the true position of the sound transmitter and the predicted position of the transmitter, with the predicted position being derived from the TDOA values. Azimuth angle in this study refers to the angle from which the waveforms arrived from, with the angle being measured from the horizontal position. In this case, one of the microphones is the reference point. Given the very close proximity of each microphone to the others as well as the relatively low sampling rate of each microphone, within 360 degrees, the algorithm was only capable of quantizing the azimuth angles to one of twelve multiples of 30 degrees. Due to this, a RandomForest model was utilized to narrow down the azimuth angle by feeding the RandomForest model the two estimated TDOA values produced by the simulation, formatted as a two-dimensional vector, as well as the ratio between the two TDOA values. Essentially, the RandomForest model acts as a classification model, classifying the azimuth angle into one of the potential azimuth angle bins, each one being one of the twelve multiples of thirty. To train and validate the RandomForest model, 24000 synthetic data points were generated using the simulation, with 80\% of these data points being delegated as training data and 20\% being delegated as validation data. A highly similar pipeline was used with an alternative multi-layer perceptron model. However, this model used deep learning and regression to find the azimuth angle as opposed to classification. Each of these data points were generated by randomly and uniformly setting the position of the transmitter in the simulation in several different places within a circle with a hundred meter radius. For each data point, the two estimated TDOA values as well as the azimuth angle were stored for training or validation purposes. In order to ensure that the models did not weigh any of the estimated TDOA values over the other, the StandardScalar class in python was applied to both TDOA values to standardized their mean and standard deviation. 

\subsection{Real World Pipeline}
On our testbed, we set up three (branded) microphones in an array, with each pair of microphones 10 centimeters away from each other, arranged in a equilateral triangle shape. After ensuring all microphones are synchronized, we use the GCC-PHAT algorithm to compute the time difference. This results in the raw, estimated TDOAs. 

The MLP model is then used to clean the TDOA and determine the azimuth angle of the transmitter source relative to the base microphone using the two raw TDOA values as well as the ratio between them. 

$$
\text{Error}(x, y) = 
(d_2 - d_1 - \Delta d_{2-1})^2 + (d_3 - d_1 - \Delta d_{3-1})^2
$$

$d_i = \sqrt{(x - x_i)^2 + (y - y_i)^2}$

$\Delta d_{2-1} = v \cdot \Delta t_{2-1}$, etc.

The above equation finds the best source location that fits the observed TDOAs. From the source location position estimate, the angle of the source can be derived. 

\subsection{Audio Filtration}
LocaGen provides a methodology to increase accuracy by removing real-world noise and factors of inconsistency. Naturally, environmental noise and predominantly that from the drone would cause high error if not filtered out. Thus, a neural network filter was trained and used in tandem with denoising software to remove over 70\% of background noise. The decision to incorporate the neural network was based on several factors, the first being the necessity of restoring distinct phases and peaks for the utilization of TDOA, something made difficult or impossible with traditional upsampling. The neural network was also incorporated to prevent distortions common in quantized signals.

The denoiser model was trained on 800 ~2.25 second audio samples featuring various noises including instruments, screaming, and random sounds. These "clean" samples were merged with drone sounds to form "mixed" ones, the objective being to remove excess noise and output an accurate clean sound wave. The one-dimensional convolutional neural network was trained with an AdamW optimizer for weight adjustment and scheduler to reduce training plateaus. Loss was defined as 20\% SI-SNR to preserve perception quality and 80\% Mean Absolute Error (MAE) to match the output to clean waveform.

The audio filtration neural network was highly successful and removed 68.80\% of unwanted noise. When analyzed on both waveform and Fast Fourier Transform graphs (shown below), large unwanted amplitudes and especially those at high frequencies were removed to yield a clean product. 

\begin{figure}[h!]
    \centering
    \includegraphics[width=0.35\textwidth]{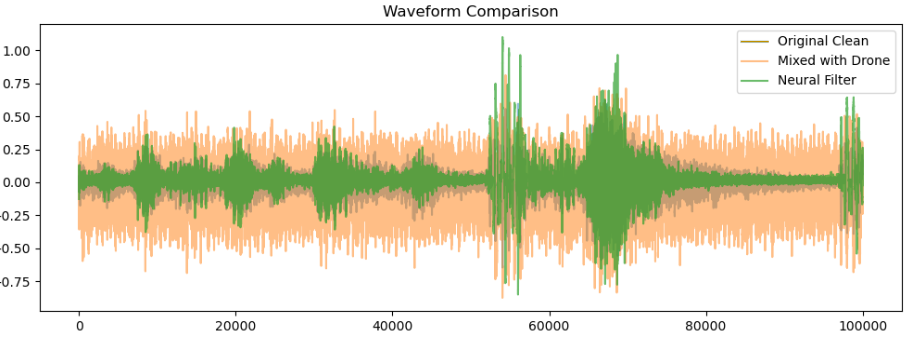}
    \caption{\textbf{Waveform Comparison} \\
    The waveform comparison for mixed noise, target noise, and filtered noise is shown. The filtered (green) and target (dark brown) waves are nearly identical, indicating success in preserving the clean form and not over-extracting data. The model also removed the majority of drone noise from the mixed input (orange), showing that noisy data was taken out.}
    \Description{The waveform comparison for mixed noise, target noise, and filtered noise is shown. The filtered (green) and target (dark brown) waves are nearly identical, indicating success in preserving the clean form and not over-extracting data. The model also removed the majority of drone noise from the mixed input (orange), showing that noisy data was taken out.}
    \label{fig:wav-diagram}
\end{figure}

\begin{figure}[h!]
    \centering
    \includegraphics[width=0.35\textwidth]{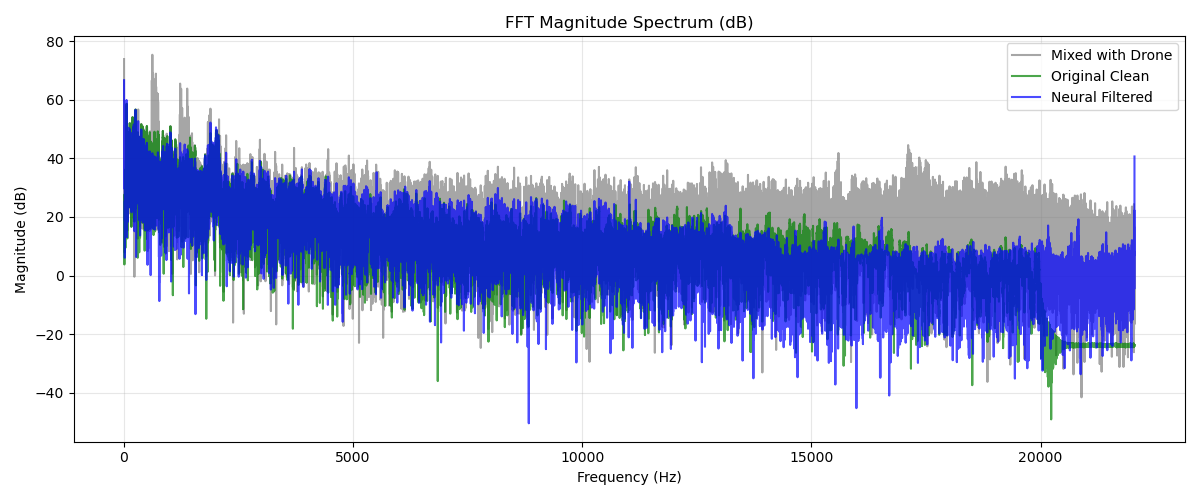}
    \caption{\textbf{FFT Diagram} \\
    The diagram shows that higher decibels (y-axis) of drone sounds which are removed at the higher frequencies (x-axis). The filtered product conforms to the target in terms of frequency and amplitude.}
    \Description{The diagram shows that higher decibels (y-axis) of drone sounds which are removed at the higher frequencies (x-axis). The filtered product conforms to the target in terms of frequency and amplitude.}
    \label{fig:fft-diagram}
\end{figure}

\section{Evaluation}

\subsection{Synthetic Approach}

It was investigated whether offsetting the sampling of the microphones would meaningfully reduce the quantization error. To this end, the simulation was utilized again to generate 10000 synthetic data points. Each data point had each of the two non-reference microphones have their sampling periods offset by a certain amount, with the sampling period of the base microphone being used as the reference point. Additionally, for each data point, the transmitter source was randomly and uniquely placed within a circle with 100 meter radius. For each data point, the error was calculated by measuring the Euclidean distance between the predicted transmitter position and the true transmitter position. Following these results, the entire dataset was moved to RStudio data analytics software. Once moved to this software, a test known as an Analysis of Variance (ANOVA) test was conducted, which is a test typically conducted to determine if, in a situation concerning at least three groups of data, there exists statistically significant differences between the mean value of each group.

Following this, more synthetic data was generated by the simulation to test the RandomForest model's performance in predicting the correct azimuth angle. The synthetic data was created by randomly placing the transmitter at different positions within a 100 meter radius circle. The transmitter was placed at different azimuth angles from 0 degrees to 360 degrees from the base microphone, incrementing the angle by 0.1 degrees.

Various types of neural networks for both regression and classification were also tested in estimating the azimuth angle. While the RandomForest model was limited to placing a prediction into different bins, a deep regression model would allow for any angle to be predicted. Thus, the RF model, a SIREN model, a multi-layer perceptron (MLP) model, and an MLP with prior Fast Fourier Transform (FFT) filtration were all trained and tested on the same dataset.

\subsection{Results}

For the ANOVA test, the aim was to determine if there was a statistically significant correlation between the two sampling offsets in each non-base microphone as well as the error function. The result of the test was a p-value of 0.266 for one of the sampling offsets and the error and another p-value of 0.185 for the other sampling offset and the error. Since a p-value greater than 0.05 indicates no statistical correlation, the study determined that there was no statistical improvement in accuracy by offsetting the sampling periods of each microphone. 

\begin{figure}[ht!]
    \centering
    \includegraphics[width=0.35\textwidth]{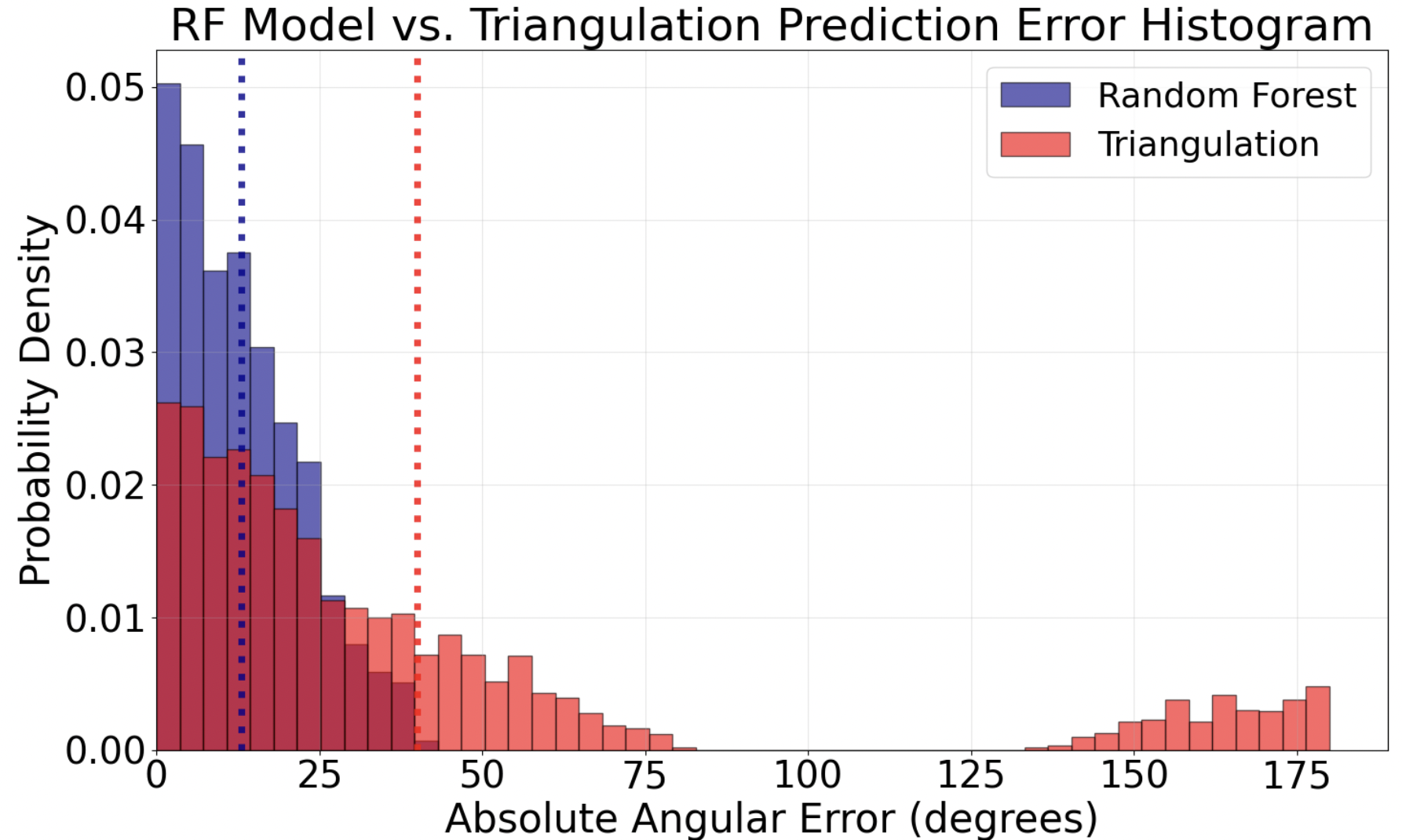}
    \caption{\textbf{RF and Triangulation Algorithm Error Comparison} \\
    This histogram compares the the error of the RF model against the error of a traditional triangulation algorithm. The distribution of absolute angle error is more concentrated around $0^\circ$ for the RF model than the algorithmic alternative, indicating great improvement with machine learning.}
    \Description{This histogram compares the the error of the RF model against the error of a traditional triangulation algorithm. The distribution of absolute angle error is more concentrated around $0^\circ$ for the RF model than the algorithmic alternative, indicating great improvement with machine learning.}
    \label{fig:error-diagram}
\end{figure}
 
The result of the testing of the RandomForest model's ability to classify the azimuth angle into one of the twelve angle bins was a mean absolute error (MAE) of 13.16 degrees, which is very close to the absolute minimum error possible of 15 degrees. Thus, it can be said that the RandomForest model has a relatively high accuracy when it comes to predict the azimuth angle and thus localizing the beam source.  

The SIREN model was completely inaccurate and ultimately had an error of about 90 degrees, making it useless. The two MLP models, however, were highly accurate and marked an improvement from the RandomForest. Due to the deep regression capabilities of the models, they yielded an MAE of 13.38 and 13.37 degrees, respectively. Even a tuned RF model which contained 24 bins instead of 12 reached a minimum MAE of 14.36 degrees.

When scaled up to sampling at 48kHz, the MLP model had an MAE of 2.87 degrees, indicating high improvement both in the ability of the model over normal triangulation and how valuable high sampling truly is.

\begin{figure}[h!]
    \centering
    \includegraphics[width=0.4\textwidth]{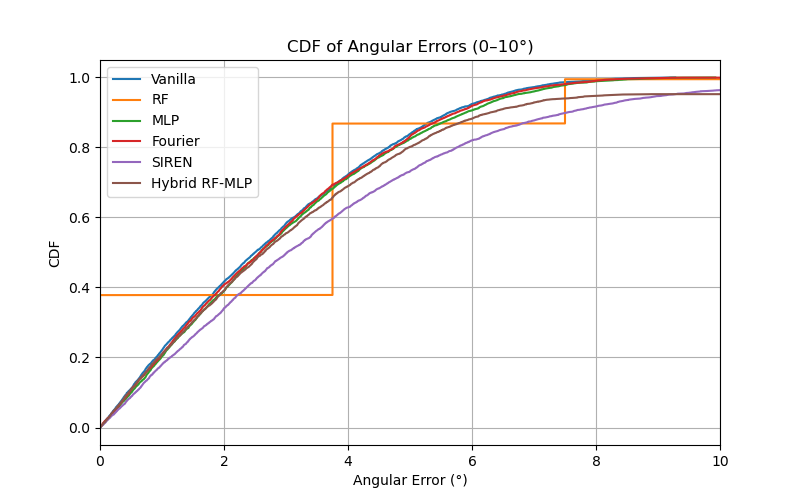}
    \caption{\textbf{Cumulative Distribution of Angle Error at 48,000 kHz} \\
    This CDF (Cumulative Distribution Plot) compares various methods of localization at 48,000 kHz. Steeper plots (such as MLP or RF) indicate more error accumulating at lower values and therefore higher accuracy.}
    \Description{This CDF (Cumulative Distribution Plot) compares various methods of localization at 48,000 kHz. Steeper plots (such as MLP or RF) indicate more error accumulating at lower values and therefore higher accuracy.}
    \label{fig:cd-diagram}
\end{figure}

\subsection{Proposed Experimental Setup}

For future collection of real-world data, GPS-coordinate data and lasers could be used to measure the precise angles and distances between the beam source and the localization microphone array. The simulation software to generate synthetic data was written in Python. 

The performance of the system should be compared against other competing localization algorithms and techniques. Our machine learning model was judged based on the total wall clock time needed to estimate the precise angle and distance. To further verify the accuracy of the model, the angle differential and distance differential should be calculated between synthetic data and real-world samples and tests.


\section{Related Work}

\subsection{TDOA and GCC-PHAT Based 2-D Plane Localization}
Several prior research attempts have been made to develop a localization method using machine learning as well as synthetic data generation. 

For instance, Lim et al. utilized two microphone arrays on two separate drones, allowing them to calculate the azimuth angle as well as the elevation of the transmitter source. Furthermore, they employed a sound filter for the drone noise, thus eliminating drone noise from the analog audio samples they received. When they tested their proposed approach in both simulated and real settings that had high signal to noise ratio (SNR), they achieved an increase in accuracy of 2.13 meters compared to traditional GCC\_Phat usage. Certain limitations associated with their study include not addressing implicit quantization errors in TDOA calculations, thus affecting localization accuracy, as well as high costs associated with two microphone arrays and two drones being required \cite{lim2025performance}. Smaller microphone arrays have been explored as well. A study conducted by Tokgoz and Panahi employed an array of three microphones that was L-shaped with the purpose of localizing sources of speech through hearing aids. Their approach employed Randomized Singular Value Decomposition (RSVD) in low SNR conditions to estimate TDOA values, yielding them a 20\% accuracy increase in simulations as well as a 25\% accuracy increase on Google Pixel recordings. However, this approach does not address the TDOA quantization problem \cite{tokgoz2021robust}. 

\subsection{Machine Learning on Synthetic Data for 2D Localization}

Increasingly, machine learning (ML) models have been employed to localize transmitter positions on a 2D plane. Jo et al., with the goal of isolating human excitatory speech while suppressing background noise to facilitate human-robot interactions, took in multichannel audio input through the microphone array on the robot. Then, Linear Prediction Residual (LPR) was applied to the input to pull out only the Excitation Source Information (ESI) while leaving out other components, with GCC-PHAT being employed after this to calculate the TDOAs. Finally, a multi-channel convolutional neural network (CNN) was employed, with each channel processing the GCC-PHAT calculation of TDOA for one pair of microphones, to classify the azimuthal angle of the transmitter source relative to the base microphone in one of several angle “bins,” which are potential angles that the azimuthal angle could be. This CNN, however, was trained on a very large ETRI-SSL dataset, which was collected by the microphone array indoors and was much harder to obtain than synthesized data \cite{jo2025sound}. Diaz-Guerra et al. also utilized an ML model for localization. They proposed \textit{Cross3D}, which essential was a CNN that was trained to localized a transmitter on a 3D plane after being trained on an extensive synthetic dataset that resembles indoor waveforms, which was constructed using Steered Response Power (SRP)-PHAT. Even with a lot of background noise present in the audio samples, the study performed highly accurately at identifying the DOA of the waveforms and, as a result, the direction of the transmitter relative to the base microphone. Notably, however, the authors deemed it ineffective to remove the TDOA calculation error that was caused as a result of the excess background noise \cite{cho2023sr}.

\subsection{Quantization Error of Small Microphone Arrays}

A study conducted by Sun et al. employed a microphone array with two microphones that were separated by only 13.66 cm and each had a sampling rate of 44.1 kHz. As a result, the time resolution is only 23 microseconds, and there were only around 35 DOA bins across 180 degrees, meaning each bin represented only around 5.14 degrees. This causes a phenomenon the study calls a "conversion error" where different DOAs that lead to the same TDOA bin cannot be resolved. \cite{zhu2019hyperear}

Despite these advancements of using TDOA estimations to localize transmitter locations, improvements are still required in resolving the quantization errors in TDOA calculations as well as rapidly and efficiently datasets to train and test localization algorithms and models. 

\section{Conclusion}

The LocaGen system presents an approach to minimizing the localization error of an audio beam using synthetic data to create a machine-learning powered corrector. The use of the simulation and synthetic data generation allows for real-world microphone arrays to have a smaller footprint while maintaining or exceeding MAE localization error levels for both angle and distance. With the development of this work, real-world SAR drone operations with the task of localizing distress signals can have smaller and lighter microphone arrays while increasing the accuracy of the source of the beam. With extensive testing in real-world conditions and the collection of the audio beam dataset, this framework can be extended for a vast array of conditions and setups. 

\balance
\bibliographystyle{IEEEtran}
\bibliography{citations.bib}

@article{lim2025performance,
  title={Performance Enhancement of Drone Acoustic Source Localization Through Distributed Microphone Arrays},
  author={Lim, Jaejun and Joo, Jaehan and Kim, Suk Chan},
  journal={Sensors},
  volume={25},
  number={6},
  pages={1928},
  year={2025},
  publisher={MDPI}
}

@article{tokgoz2021robust,
  title={Robust three-microphone speech source localization using randomized singular value decomposition},
  author={Tokgoz, Serkan and Panahi, Issa MS},
  journal={IEEE Access},
  volume={9},
  pages={157800--157811},
  year={2021},
  publisher={IEEE}
}

@article{jo2025sound,
  title={Sound Source Localization Using Deep Learning for Human--Robot Interaction Under Intelligent Robot Environments},
  author={Jo, Hong-Min and Kim, Tae-Wan and Kwak, Keun-Chang},
  journal={Electronics},
  volume={14},
  number={5},
  pages={1043},
  year={2025},
  publisher={MDPI}
}

@article{concretePaper,
  title={TDOA-based localization of cracking sound events with minimal-error microphone subsets},
  author={Kocur, Georg Karl and Kumar, Bharath and Markert, Bernd},
  journal={NDT \& E International},
  volume={147},
  number={103211},
  year={2024},
  publisher={ScienceDirect}
}

@article{ICUPaper,
  title={Acoustic source localization with microphone arrays for remote noise monitoring in an Intensive Care Unit},
  author={Müller-Trapet, Markus and Cheer, Jordan and Fazi, Filippo Maria and Darbyshire, Julie and Young, J Duncan},
  journal={Appl Accoust.},
  volume={139},
  number={93},
  year={2018},
  publisher={PubMed Central}
}

@misc{costSource,
  author       = {Jouav Unmanned Aircraft Systems},
  title        = {Best Search and Rescue (SAR) Drones},
  howpublished = {\url{https://www.jouav.com/search-and-rescue-drone}},
}

@misc{haitiEarthquake,
  author       = {International Commission on Missing Persons},
  title        = {Haiti},
  howpublished = {\url{https://icmp.int/the-missing/where-are-the-missing/haiti/}},
}

@inproceedings{cho2023sr,
  title={SR-SRP: Super-resolution based SRP-PHAT for sound source localization and tracking},
  author={Cho, Jae-Heung and Chang, Joon-Hyuk},
  booktitle={Proc. Interspeech 2023},
  pages={3769--3773},
  year={2023}
}

@inproceedings{zhu2019hyperear,
  title={Hyperear: Indoor remote object finding with a single phone},
  author={Zhu, Hongzi and Zhang, Yuxiao and Liu, Zifan and Chang, Shan and Chen, Yingying},
  booktitle={2019 IEEE 39th International Conference on Distributed Computing Systems (ICDCS)},
  pages={678--687},
  year={2019},
  organization={IEEE}
}


\end{document}